\documentclass[aps
]{revtex4}

\usepackage{amscd}\usepackage{amssymb}\usepackage{amsbsy}

\begin{document}

\newcommand{\R}{\mathbb{R}}
\newcommand{\C}{\mathbb{C}}
\newcommand{\proofend}{\hfill\rule{0.2cm}{0.2cm}}
\newcommand{\Cinfty}{{\mathcal C}^\infty}
\newcommand{\ve}{\varepsilon}
\newcommand{\dfrac}[2]{{\displaystyle\frac{#1}{#2}}}
\newcommand{\tfrac}[2]{{\textstyle\frac{#1}{#2}}}
\newcommand{\vp}{{\mathrm vp}}
\newcommand{\Pf}{\mbox{Pf}}
\newcommand{\signum}{\mbox{signum}}
\newenvironment{lessimportant}{\begingroup\small}{\endgroup}
\newenvironment{evenlessimportant}{\begingroup\footnotesize}{\endgroup}
\newenvironment{moreimportant}{\begingroup\large}{\endgroup}
\newcommand{\normal}{\rm}
\newcommand{\emphaz}{\it}
\newcommand{\varemphaz}{\bf}   
\newcommand{\Cc}{{\cal C}}
\newcommand{\todo}[1]{$\clubsuit$\ {\tt #1}\ $\clubsuit$}  
\newcommand{\beas}{\begin{eqnarray*}}\newcommand{\eeas}{\end{eqnarray*}}
\newcommand{\esm}{\ensuremath{{\mathcal E}_M} }\newcommand{\eps}{\varepsilon} 
\newcommand{\es}{\ensuremath{{\mathcal E}} }\newcommand{\N}{\mathbb N} 
\newcommand{\nn}{\nonumber}\newcommand{\ns}{\ensuremath{{\mathcal N}} } 
\newcommand{\T}{{\cal T}}\newcommand{\gs}{\ensuremath{{\mathcal G}} } 
\newcommand{\cl}{\mbox{\rm cl}}\newcommand{\al}{\alpha}      
\newcommand{\supp}{\mathop{\mathrm{supp}}}
%
%
%
%
%
%
%

\title[Distributional Schwarzschild geometry]{Remarks on the distributional Schwarzschild geometry}
\author{J. Mark Heinzle}
\email{mheinzle@galileo.thp.univie.ac.at}
\affiliation{Institute for Theoretical Physics, University of Vienna, Boltzmanng.\ 5, A-1090 Wien, Austria}
\author{Roland Steinbauer}
\email{roland.steinbauer@univie.ac.at}
\affiliation{Department of Mathematics, University of Vienna, Strudlhofg.\ 4, A-1090 Wien, Austria}
\affiliation{Institute for Theoretical Physics, University of Vienna, Boltzmanng.\ 5, A-1090 Wien, Austria}

\begin{abstract}
\qquad\vskip18pt
\noindent
This work is devoted to a mathematical analysis of the distributional
Schwarzschild geometry. The Schwarzschild solution is extended to
include the singularity; the energy momentum tensor becomes a $\delta$-distribution
supported at $r=0$. Using generalized distributional
geometry in the sense of Colombeau's (special) construction the nonlinearities are treated in a 
mathematically rigorous way. Moreover, generalized function techniques are used 
as a tool to give a unified discussion of various approaches taken in the literature so far;
in particular we comment on geometrical issues.  
\vskip12pt
\noindent
{\bf Physics and Astronomy Classification Scheme (2001):} 
Primary: 04.20.Cv; 
secondary: 04.20.Jb, 
04.20.Dw, 
02.30.Sa, 
02.40.Vh.
\end{abstract}
\maketitle

\section {Introduction}\label{introduction} 
Since the formulation of the first singularity theorems it is generally conceded that 
singular spacetimes are of fundamental importance in general relativity.
Geometrically, a singularity is defined via the notion of (geodesic) incompleteness,
a viewpoint which fits in the singularity theorems of Hawking and Penrose 
(see, e.g., \cite{hawking+ellis}, Chap.\ 8), forcing us to regard a singularity as 
some kind of singular boundary point of spacetime. 
Recently, as an alternative, it has been suggested to describe (mild) singularities as 
internal points, where the field equations are satisfied in a weak (probably distributional)
sense (cf.\ \cite{clarkeINDIEN}). General relativity as a physical theory is governed by 
particular physical equations; the focus of interest is the breakdown of physics which 
need not coincide with the breakdown of geometry.

In the context of conical spacetimes algebras of generalized functions \cite{colombeau1, colombeau2} 
have been used to overcome the problem of simultaneously dealing 
with singular (i.e., distributional) metrics and the nonlinearities of general 
relativity \cite{cvw,wilson,invc}. These techniques allow to assign to the cone metric a 
singular energy momentum tensor supported on a submanifold of codimension two, 
which, by a result of Geroch and Traschen \cite{gt}, is not possible within classical (i.e., linear)
distribution theory.

The main focus of this work is a (nonlinear) distributional description of the
Schwarz\-schild spacetime. Although the nature of the Schwarzschild singularity is much ``worse'' than 
the quasi-regular conical singularity, there are 
several distributional treatments in the 
literature 
(\cite{Balasin/Nachbagauer:1993,Balasin/Nachbagauer:1994,Kawai/Sakane:1997,Pantoja/Rago:1997,
Pantoja/Rago:2000}), mainly motivated by the following considerations: 
the physical interpretation of the Schwarz\-schild 
metric is clear as long as we consider it merely as an exterior (vacuum) solution 
of an extended (sufficiently large) massive spherically symmetric body. Together with 
the interior solution it describes the entire spacetime. The concept of point 
particles---well understood in the context of linear field theories---suggests a
mathematical idealization of the underlying physics: one would like to view the 
Schwarz\-schild solution as defined on the entire spacetime and regard it as 
generated by a point mass located at the origin and acting as the 
gravitational source. This of course amounts to the question of whether one can 
reasonably ascribe distributional curvature quantities to the Schwarzschild 
singularity at the origin.

The emphasis of the present work lies on mathematical rigor. We derive
the ``physically expected'' result for the distributional energy momentum 
tensor of the Schwarzschild geometry, i.e., $T^0_{\:0}=8\pi m\delta^{(3)}(\vec{x})$, in a 
conceptually satisfactory way. Additionally, we set up
a unified language to comment 
on the respective merits of some of the approaches taken so far.
In particular, we discuss questions of differentiable structure as
well as smoothness and degeneracy problems of the regularized metrics, and present possible
refinements and workarounds. These aims are accomplished using the 
framework of nonlinear generalized functions (Colombeau algebras) \cite{colombeau1,
colombeau2} and, in particular, the geometric approach taken in \cite{ndg,gcf}.
\vskip12pt

The paper is organized in the following way: in section 
\ref{prereq} we discuss the conceptual as well as the mathematical
prerequisites. In particular we comment on
geometrical matters (differentiable structure, coordinate invariance)
and recall the basic facts of nonlinear distributional geometry
in the context of algebras of generalized functions. Moreover, we
derive sensible regularizations of the singular functions to be used 
throughout the paper. Section \ref{afirstapproachtotheproblem} is devoted to a first 
approach to the problem;
a detailed discussion follows in section \ref{com+prob}: 
we comment on problems and obstacles associated with 
the direct approach.
Finally, in section \ref{thekerrschildapproach}, we present a new 
conceptually satisfactory method to derive the main result. 
Overly technical calculations are shifted to various appendices.
In the final section \ref{thehorizon} we investigate the horizon
and describe its distributional curvature. 
Using nonlinear distributional geometry and generalized functions
it seems possible to show that the horizon singularity is only
a coordinate singularity {\em without leaving Schwarzschild coordinates}.

\setcounter{equation}{0}\section {Prerequisites}\label{prereq} 

To begin with, let us have a look at the conceptually much simpler problem
of point charges in Maxwell's theory and consider the Coulomb solution 
$\frac{1}{r}$ of an extended spherically symmetric body. 
In an idealized picture the charged body is reduced to a point charge, and 
this way of looking at the problem has proven to be very fruitful, 
mainly due to the following two reasons: first, the function 
$\frac{1}{r}\in\Cc^2(\R^3\backslash\{0\})$, since also in $L^1_{loc}(\R^3)$, naturally gives rise to
a distribution on $\R^3$. Reinserting this distributional potential into the field equation
we obtain $\Delta \frac{1}{r}=-4\pi \delta\,$, which has the clear physical interpretation as
the charge density of a point charge.
Second, also in accordance with physical intuition, the situation may be interpreted 
in terms of
the following sensible regularization scenario: 
consider a regularization of the ``singular'' potential by any sequence of (say smooth) functions
converging weakly to $\frac{1}{r}$. Then, by virtue of linearity
of the field equation, distribution theory guarantees that the corresponding 
sequence of charge
densities will converge weakly to $-4\pi \delta$, i.e., the density of the point charge.

The general relativistic case is much more involved. 
Consider the Schwarzschild metric inside the horizon: 
extending the spacetime 
to $r=0$ we are confronted with several distinct problems. First---according to 
the standard picture of general relativity---no manifold structure is
given at the singularity $r=0$, since the field equations are meaningless there within 
the smooth category. In addition, the differentiable structure of the extended manifold cannot 
be uniquely determined from the differentiable structure of the original spacetime.
This problem is dealt with in the relevant literature by fixing some differentiable
structure by hand, most often the one induced by Cartesians associated to 
Schwarzschild coordinates. 

In analogy to the Maxwell case, we want to regard
the metric as a distribution on the whole extended spacetime. Now, the
second conceptual problem is due to the inherently nonlinear nature
of general relativity: no distributional meaning can be given to the field equations, since 
it is not possible to calculate the curvature from a distributional metric.
In the literature, this obstacle is circumvented by using various---more or 
less---ad-hoc regularization approaches in order to calculate a regularized
Ricci tensor within the smooth category. Eventually, its distributional 
limit is computed and---via the field equations---a distributional energy
momentum tensor is obtained. This tensor may then be interpreted as 
distributional source of the Schwarzschild geometry
\cite{Balasin/Nachbagauer:1993,
Kawai/Sakane:1997,Pantoja/Rago:1997,Pantoja/Rago:2000,Balasin/Nachbagauer:1994}.
However, using ad-hoc regularizations we are confronted with
the problem of regularization independence of the results which
may not be suitably addressed within this setting.

In this work, while arguing form a related point of view, we are going to use a
different apparatus to deal with the nonlinearities of GR: 
the theory of algebras of generalized functions 
gives us the additional flexibility and power of a rigorous 
mathematical framework in which distributions may undergo nonlinear operations. 
In particular, following the procedure of \cite{cvw}, we will first model the distributionally 
extended Schwarzschild metric by a generalized metric obtained by a suitable (and general) 
regularization procedure. 
Then, after entering the generalized framework (cf.\ \cite{gcf}) we may calculate all the 
relevant curvature quantities from the generalized metric and subject it to the field equations. 
Note that within the generalized setting the field equations possess a well defined meaning. 
Finally, we may descend to the distributional level for the purpose of interpretation 
using the concept of association (see below).

Note that, in general, a regularization procedure depends on the coordinate system
it is performed in (for a diffeomorphism invariant notion of regularization using paths of
mollifiers see \cite{Grosser:2000,adv}). 
However, since we had to fix the differentiable structure beforehand
this is no further restriction. 
Actually, geometric considerations play an important role:
as shown below, a brute force regularization 
attempt does not lead to a sensible description of the problem at hand. 
Indeed, we shall see
that a satisfactory treatment of the distributional Schwarzschild spacetime 
has to use the Kerr-Schild form of the Schwarzschild metric (which fixes both
the differentiable structure and the coordinates); moreover, it must be retained 
during the whole regularization process. Note that this is in accordance with physical
intuition since in the Kerr-Schild form the radial coordinate retains its
spacial character near the singularity which of course is not the case in Schwarzschild
coordinates.

In the remainder of this section we are going to introduce some mathematical prerequisites.
First, we are going to shortly recall generalized 
tensor analysis and generalized curvature (in Colombeau's so-called special setting).
For all further details we refer the reader to \cite{ndg,gcf}. Second, 
we explicitly calculate the
regularization of the relevant components of the metric tensor to be used throughout the paper.

\subsection*{Nonlinear distributional geometry}

The basic idea of Colombeau's theory of generalized functions \cite{colombeau1,colombeau2}
is regularization by sequences (nets) of smooth functions and the use of asymptotic estimates in terms of
a regularization parameter $\varepsilon$. Let  $(u_\eps)_{\eps\in(0,1]}$
with $u_\eps\in\Cc^\infty(M)$ for all $\eps$ ($M$ a separable, smooth orientable Hausdorff manifold 
of dimension $n$). The algebra of generalized 
functions on $M$ is defined as the quotient $\gs(M):=\esm(M)/\ns(M)$
of the space $\esm(M)$ of sequences of moderate growth modulo the space $\ns(M)$ of negligible
sequences. More precisely the notions of moderateness resp.\ negligibility are defined by the following
asymptotic estimates (${\mathfrak X}(M)$ denoting the space of smooth vector fields on $M$)
\beas
        \esm(M)&:=&\{ (u_\eps)_\eps:\ \forall K\subset\subset M,\,
             \forall k\in\N_0\,\exists N\in\N\ \\
             &&\forall \xi_1,\dots,\xi_k\in{\mathfrak X}(M):\ 
             \sup_{p\in K}|L_{\xi_1}\dots L_{\xi_k}\,u_\eps(p)|=O(\eps^{-N})\} 
      \\
        \ns(M)&:=& \{ (u_\eps)_\eps:\ \forall K\subset\subset M,\,\forall k,q \in\N_0\\
        &&\forall \xi_1,\dots,\xi_k\in{\mathfrak X}(M):\ 
        \sup_{p\in K}|L_{\xi_1}\dots L_{\xi_k}\,u_\eps(p)|=O(\eps^{q}))\}
\eeas

Elements of $\gs(M)$ are denoted by $u=\cl[(u_\eps)_\eps]=
(u_\eps)_\eps+\ns(M)$. With componentwise operations $\gs(M)$ is a fine sheaf of 
differential algebras with respect to the Lie derivative defined by 
$L_\xi u:=\cl[(L_\xi u_\eps)_\eps]$. The spaces of moderate resp.\ negligible sequences 
and hence the algebra itself may be characterized locally, i.e., $u\in\gs(M)$ 
iff $u\circ\psi_\al\in\gs(\psi_\al(V_\al))$ for all charts
$(V_\al,\psi_\al)$, where on the open set $\psi_\al(V_\al)\subset\R^n$ in the respective
estimates Lie derivatives are replaced by partial derivatives.
Smooth functions are embedded into $\gs$ simply by the ``constant'' embedding $\sigma$, i.e.,
$\sigma(f):=\cl[(f)_\eps]$, hence $\Cinfty(M)$ is a faithful subalgebra of $\gs(M)$.
On open sets of $\R^n$ compactly supported distributions are embedded into $\gs$ via convolution with
a mollifier $\rho\in{\mathcal S}(\R^n)$ with unit integral satisfying $\int\rho(x)x^\alpha dx=0$ for
all $|\alpha|\geq 1$; more precisely setting $\rho_\eps(x)=(1/\eps^n)\rho(x/\eps)$ we
have $\iota(w):=\cl[(w*\rho_\eps)_\eps]$. In case $\supp(w)$ is not compact one uses 
a sheaf-theoretical construction. However, in the special case of the  
functions to be treated in the context of the Schwarzschild metric this will not be necessary
(see below). From the explicit formula it is clear that (on open subsets
of Euclidean space) embedding commutes with partial differentiation. 
On a general manifold, however, there is no canonical embedding of ${\mathcal D}'$ available; a suitable
replacement (cf.\ \cite{ndg}) is provided by physically motivated modeling and the use
of the notion of association (see below). Inserting $p\in M$ into $u\in\gs(M)$ yields a well 
defined element of the ring of constants (also called generalized numbers) 
${\cal K}$ (corresponding to ${\mathbb K}=\R$ 
resp.\ $\C$), defined as the set of moderate nets of numbers ($(r_\eps)_\eps \in{\mathbb K}^{(0,1]}$ with
$|r_\eps| = O(\eps^{-N})$ for some $N$) modulo negligible nets
($|r_\eps| = O(\eps^{m})$ for each $m$).
Finally, generalized functions on $M$ are
characterized by their generalized point values, i.e., by their values on points
in $\tilde M_c$, the space of equivalence classes of compactly supported nets $(p_\eps)_\eps
\in M^{(0,1]}$ with respect to the relation $p_\eps\sim p'_\eps:\Leftrightarrow 
d_h(p_\eps,p'_\eps)=O(\eps^m)$ for all $m$, where $d_h$ denotes the distance on $M$ induced 
by any Riemannian metric.

The $\gs(M)$-module of generalized sections in vector bundles---especially the space of 
generalized tensor fields $\gs^r_s(M)$---is defined along the same lines using analogous 
asymptotic estimates with respect to the norm induced by any Riemannian metric on the respective 
fibers. However, it is more convenient to use the following algebraic description of generalized
tensor fields
\begin{equation}\label{tensorp}
  \gs^r_s(M)=\gs(M)\otimes{\mathcal T}^r_s(M)\,,
\end{equation}
where ${\mathcal T}^r_s(M)$ denotes the space of smooth tensor fields and the tensor product is
taken over the module $\Cinfty(M)$. Hence generalized tensor fields are just given by classical ones
with generalized coefficient functions. Many concepts of classical tensor analysis carry over to the
generalized setting \cite{ndg}, in particular Lie derivatives with respect to 
both classical and generalized vector fields, Lie brackets, exterior algebra, etc. 
Moreover, generalized tensor fields may also be viewed as
$\gs(M)$-multilinear maps taking generalized vector and covector fields to generalized functions, 
i.e., as $\gs(M)$-modules we have
\[
   \gs^r_s(M)\cong L_{\gs(M)}(\gs^0_1(M)^r,\gs^1_0(M)^s;\gs(M)).
\]
In particular a generalized metric is defined to be a symmetric, generalized $(0,2)$-tensor field
$ g_{ab}=\cl[( g_{ab\,\eps})_\eps]$ (with its index independent of $\eps$ and) whose 
determinant $\det( g_{ab})$ is invertible 
in $\gs(M)$. The latter condition is equivalent to the following notion called strictly nonzero 
on compact sets: for any representative $\det( g_{ab\,\eps})_\eps$ of
$\det( g_{ab})$ we have $\forall K\subset\subset M\ \exists m\in\N:\ \inf_{p\in K}
|\det( g_{ab\,\eps})|\geq\eps^m$ for all $\eps$ small enough. This notion 
captures the intuitive idea of a generalized metric to be a sequence of classical metrics
approaching a singular limit in the following sense: $ g_{ab}$ is a generalized
metric iff (on every relatively compact open subset $V$ of $M$) there exists a representative 
$( g_{ab\,\eps})_\eps$ of $ g_{ab}$ such that for fixed $\eps$ (small enough) $ g_{ab\,\eps}$
(resp.\ $ g_{ab\,\eps}|_V$) is a classical pseudo-Riemannian metric and $\det( g_{ab})$
is invertible in the algebra of generalized functions. A generalized
metric induces a $\gs(M)$-linear isomorphism from $\gs^1_0(M)$ to $\gs^0_1(M)$ and the inverse
metric $ g^{ab}:=\cl[( g_{ab\,\eps}^{-1})_\eps]$ is a well defined element of $\gs^2_0(M)$
(i.e., independent of the representative $( g_{ab\,\eps})_\eps$).
Also the generalized Levi-Civita connection as well as the generalized Riemann-, Ricci- and 
Einstein tensor of a generalized metric are defined simply by the usual coordinate formulae
on the level of representatives.
 
Finally, the setting introduced above displays maximal consistency (in the light of L.\ Schwartz 
impossibilty result \cite{simp}) with respect to smooth resp.\ distributional geometry
most conveniently formalized in terms of the notion of association. A generalized 
function $u\in\gs(M)$ is
called associated to zero, $u\approx 0$, if one (hence any) representative $(u_\eps)_\eps$ converges 
to zero weakly. (In a sloppy fashion we shall often write $u_\eps\approx 0$.)
The equivalence relation $u\approx v:\Leftrightarrow u-v\approx0$ gives rise
to a linear quotient of $\gs$ that extends distributional equality. Moreover
we call a distribution $w\in{\mathcal D}'(M)$ the distributional shadow or macroscopic aspect of
$u$ and write $u\approx w$ if for all compactly supported $n$-forms $\nu$ and one (hence any) 
representative $(u_\eps)_\eps$
\[\lim_{\eps\to 0}\int\limits_Mu_\eps\nu=\langle w,\nu\rangle,\]
where $\langle\,,\,\rangle$ denotes the distributional action. By (\ref{tensorp}) the concept 
of association extends to generalized tensor fields in a natural way.

\subsection*{Regularizations of the singular functions
occurring in the Schwarzschild metric}
The two most important singular functions we will work with throughout this paper
(namely the singular components of the Schwarzschild metric)
are $\frac{1}{r}$ and $\frac{1}{r-c}$ ($r = \|\vec{x}\|\,$; $c$ a positive constant).
Since  $\frac{1}{r}\in L^1_{loc}({\mathbb R}^3)$, it gives rise to the
regular distribution $\frac{1}{r} \in {\mathcal D}^{\prime}({\mathbb R}^3)$. 
By convolution with a mollifier $\rho$ (adapted to the symmetry
of the spacetime, thus chosen radially symmetric) 
we embed it into the Colombeau algebra ${\mathcal G}({\mathbb R}^3)$
\begin{equation}\label{embed1overr}
  \frac{1}{r}\quad\stackrel{\iota}{\rightarrow}\quad\iota(\frac{1}{r}) =
  (\frac{1}{r}\ast\rho_\varepsilon)=:
  (\frac{1}{r})_\varepsilon\,.
\end{equation}
Using radial symmetry of the convoluted function and inserting 
$\rho_\varepsilon(r) = \frac{1}{\varepsilon^3} \rho(\frac{r}{\varepsilon})\,$ 
we obtain
\begin{equation}\label{1overrepsilonformula} 
  (\frac{1}{r})_\varepsilon=
  \frac{4\pi}{r}\int\limits_0^{r/\varepsilon} dt\: t^2 \,\rho(t) +
  \frac{4\pi}{\varepsilon} \int\limits_{r/\varepsilon}^\infty dt \:t \,
  \rho(t)\,.
\end{equation}
It is easy to confirm that    
$(\frac{1}{r})_\varepsilon= \sigma(\frac{1}{r})_\varepsilon = \frac{1}{r}$ 
on ${\mathbb R}^3\backslash \{0\}$,
and at the origin we have $(\frac{1}{r})_\varepsilon \Big|_{r=0} = 
\frac{4\pi}{\varepsilon}\int_0^\infty dt \:t \rho(t)\:$.

In contrast to $\frac{1}{r}$, the function $\frac{1}{r-c}$
is not in $L^1_{loc}(\mathbb{R}^3)$. A canonical 
regularization (in the sense of Gelfand-Shilov \cite{gs}) 
is the principal value ${\mathrm vp}(\frac{1}{r-c}) \in {\mathcal D}^\prime({\mathbb R}^3)$
which can be embedded into ${\mathcal G}(\R^3)$. 
\begin{equation}\label{embed1overrminusc}
  \frac{1}{r-c} \quad\stackrel{}{\rightarrow}\quad
  {\mathrm vp}(\frac{1}{r-c})\in{\mathcal D}^\prime({\mathbb R}^3)
  \quad\stackrel{\iota}{\rightarrow}\quad\iota(\vp(\frac{1}{r-c})) =:
  (\vp(\frac{1}{r-c}))_\varepsilon\,. 
\end{equation}
Making use of ${\mathrm vp}(\frac{1}{r-c})=\frac{\partial}{\partial r} \log|r-c|$ 
we obtain $\iota(\vp(\frac{1}{r-c}))(x)=$
\[ 
(1+r\frac{\partial}{\partial r})
\int d^3y \frac{1}{|x-y|} \log |\,|x-y|\,-c| \,\rho_\varepsilon(y) - 
\frac{\partial}{\partial x^i} \int d^3y \,y^i \frac{1}{|x-y|} 
\log |\,|x-y|\,-c| \,\rho_\varepsilon(y)
\]
and finally for $v\geq c$
\begin{eqnarray}\label{vpembed}
\nonumber
\iota({\mathrm vp}(\frac{1}{r-c}))(x)  & = &   
\frac{4\pi}{r-c} \int\limits_0^{r-c} ds \rho_\varepsilon(s) s^2 +
\frac{4\pi}{r} \int\limits_{r-c}^\infty ds \rho_\varepsilon(s) s^2 + \\
\nonumber
& & 
+ \frac{4\pi}{r-c}\frac{c}{r}
\int\limits_0^{r-c} ds \rho_\varepsilon(s) s^2 \sum_{l=1}^\infty
\frac{1}{2l+1} (\frac{s}{r-c})^{2l} + \\
& &
+ \frac{4\pi}{r-c} \frac{c}{r}
\int\limits_{r-c}^\infty ds \rho_\varepsilon(s)
(r-c)^2 \sum_{l=0}^\infty \frac{1}{2l+1} (\frac{r-c}{s})^{2l}. 
\end{eqnarray}
For $0< r\leq c$ the roles of $r$ and $c$ are interchanged and at the origin
we obtain $\vp(\frac{1}{r-c})_\varepsilon\,\Big|_{r=0} = -\frac{1}{c} + O(\varepsilon)$.

\section{A first approach to the problem}\label{afirstapproachtotheproblem} 
In this section we present a first approach to the ``Schwarzschild
point mass problem'', thereby essentially following earlier treatments in the literature 
(\cite{Balasin/Nachbagauer:1993,Kawai/Sakane:1997,Pantoja/Rago:1997,Pantoja/Rago:2000}). 
However, we are going to use the language of nonlinear distributional geometry
introduced above in order to obtain a unified view, which will enable us to
carry out a detailed analysis of the previous approaches in the next section. 

In the usual Schwarzschild coordinates $(t, r>0, \theta, \phi)$ 
the metric takes the form
\begin{equation}\label{ssmetric}
ds^2 = h(r)\: dt^2 - h(r)^{-1}\: dr^2 + r^2 \:d\Omega^2
\mbox{ with } h(r)=-1 + \frac{2m}{r}.
\end{equation}
Following the above discussion  we consider 
the singular metric coefficient $h(r)$ as an element of $L^1_{loc}({\mathbb R}^3) 
\subseteq{\mathcal D}'({\mathbb R}^3)$ 
and embed it into ${\mathcal G}({\mathbb R}^3)$ by convolution with a mollifier.
Note that, accordingly, we have fixed the differentiable structure of the manifold:
the Cartesian coordinates associated with the spherical Schwarzschild coordinates in 
(\ref{ssmetric}) are extended through the origin. We have
\begin{equation}\label{hregul} 
h(r)=-1+\frac{2m}{r}
\quad\stackrel{\iota}{\rightarrow}\quad
\iota(h(r)) = 
h_\varepsilon(r) = -1 + 2m (\frac{1}{r})_\varepsilon\in {\mathcal G}({\mathbb R}^3)\,, 
\end{equation}
where $(\frac{1}{r})_\varepsilon$ is given by (\ref{1overrepsilonformula}).
Inserting (\ref{hregul}) into (\ref{ssmetric}) we obtain a generalized object
modeling the singular Schwarzschild metric, i.e.,
\begin{equation}\label{ssmetricembed} 
ds^2_\epsilon = 
h_\epsilon(r)\: dt^2 - h_\epsilon(r)^{-1}\: dr^2 
+ r^2 \:d\Omega^2\,.
\end{equation}
The generalized Ricci tensor may now be calculated componentwise using the classical formulae
\begin{eqnarray}\label{r00epsilon} 
(R^0_{\:0})_\varepsilon = (R^1_{\:1})_\varepsilon &=& \frac{1}{2} 
\left(h_\varepsilon^{\prime\prime} + \frac{2}{r} h_\varepsilon^\prime\right) =
\frac{1}{2}\: \Delta h_\varepsilon\\
\label{r22epsilon} 
(R^2_{\:2})_\varepsilon = (R^3_{\:3})_\varepsilon &=& \frac{h_\varepsilon^\prime}{r}+ 
\frac{1+h_\varepsilon}{r^2}\,.
\end{eqnarray}
Due to the linear structure of $R^0_{\:0}$ it is evident that it 
is associated to a constant times the $\delta$-distribution, i.e.,
\begin{equation}
(R^0_{\:0})_\varepsilon = \frac{1}{2}\: \Delta h_\varepsilon = 
m\:\Delta (\frac{1}{r})_\varepsilon \:\rightarrow\:
-4 \pi m \delta \quad(\varepsilon\rightarrow 0).
\end{equation}
Investigating the weak limit of the angular components of the Ricci tensor 
(using the abbreviation $\tilde{\Phi}(r) = \int \sin\theta d\theta d\phi \:\Phi(\vec{x})$)
we get (cf.\ appendix \ref{appendixA:mollifierintegrals})
\begin{eqnarray*}
\int (R^2_{\:2})_\varepsilon \Phi\: d^3x & = &
\int (r h_\varepsilon^\prime + 1 + h_\varepsilon) \tilde{\Phi}(r) dr = \\
& \stackrel{(\ref{1overrepsilonformula})}{=} &
8 \pi m \int \frac{1}{\varepsilon} \,[\int_{r/\varepsilon}^\infty t \rho(t) dt]
\:\tilde{\Phi}(r) dr =
8 \pi m \int dx \,\tilde{\Phi}(\varepsilon x) \int_x^\infty t \rho(t) dt \\
& \rightarrow & 32 \pi^2 m \,\Phi(0) \int\limits_0^\infty dx 
\int_x^\infty t \rho(t) dt\: 
\stackrel{(\ref{mollifierintegral3})}{=}\: 8 \pi m \langle\delta|\Phi\rangle
\qquad (\varepsilon\rightarrow 0).
\end{eqnarray*}
Hence, the Ricci tensor and the curvature scalar $R$ are of $\delta$-type, i.e.,
\begin{equation}\label{ricciresults}
R^0_{\:0}\,=\,R^1_{\:1} \approx -4 \pi m\,\delta \qquad
R^2_{\:2} = R^3_{\:3} \approx  8 \pi m\,\delta \qquad
R \approx  \pi m\,\delta.
\end{equation}

Equations~(\ref{ricciresults}) are obviously given
in spherical coordinates. Strictly speaking this is not sensible, because
the basis fields $\{\frac{\partial}{\partial r},
\frac{\partial}{\partial\phi}, \frac{\partial}{\partial\theta}\}$
are not globally defined. 
Representing distributions concentrated at
the origin requires a basis regular at the origin.
Transforming the results for $(R^i_{\:j})_\varepsilon$ 
(i.e., (\ref{r00epsilon}) and~(\ref{r22epsilon})) into Cartesian coordinates
associated with the spherical ones (i.e., $\{r,\theta,\phi\}\leftrightarrow\{x^i\}$) 
we obtain, e.g., for the Einstein tensor
\begin{equation}\label{einsteintensorincartesians} 
G^i_{\:j} \;\approx\; -8 \pi m\,\delta\:\delta^i_0\delta^0_j.
\end{equation}

Note that the use of the particular regularization~(\ref{hregul}) 
is not essential here.
We could have replaced~(\ref{hregul}) by any other smooth ad-hoc regularization of $h(r)$,
as has been done, e.g., in \cite{Kawai/Sakane:1997}, 
by setting $h_\varepsilon(r) = -1 + \frac{2m}{\sqrt{r^2+\varepsilon^2}}$. 
Indeed, we can show that the results (\ref{ricciresults}) 
hold for all regularizations, i.e., for all sequences
of the form $h_\varepsilon(r) = -1 + 2m s_\varepsilon(r) \rightarrow h(r)$ (i.e.,
$\forall\: s_\varepsilon$ smooth, $s_\varepsilon \rightarrow \frac{1}{r}$ in ${\mathcal D}'$).
For the $(0,0)$- and $(1,1)$-components of the Ricci tensor the result follows
from the special form of (\ref{r00epsilon}). For the angular components
(cf.\ (\ref{r22epsilon})) we write
\begin{equation}
2m \int\limits_0^\infty r^2 
(\frac{s_\varepsilon^\prime(r)}{r} + \frac{s_\varepsilon(r)}{r^2}) 
\,\tilde{\Phi}(r) \,dr=
-2m \int\limits_0^\infty dr\, r^2 s_\varepsilon\, 
\frac{1}{r}\, \tilde{\Phi}^\prime(r)
\,\rightarrow\, 8\pi m \Phi(0),
\end{equation}
where in the last step we used the fact that $\frac{1}{r}\,
\tilde{\Phi}^\prime\in{\mathcal D}(\R)$.

\section{Comments and problems}\label{com+prob}

In order to be able to calculate the curvature from the metric we must
keep the regularization $h_\varepsilon(r)$ smooth on the entire spacetime.
This fact---although somewhat hidden because we worked with spherical 
coordinates---is essential from the conceptual point of view. 
In fact, choosing a regularization $h_\varepsilon(r)$ which is smooth only 
on ${\mathbb R}^3\backslash\{0\}$ is not sufficient
to derive the result as is explicitly shown by the following counterexample. 
Set $h_\varepsilon = -1 + 2m s_\varepsilon$ and 
$s_\varepsilon = \frac{1}{r}\, o_\varepsilon$  (with
$o_\varepsilon \rightarrow 1$ weakly) consisting of
regular distributions, so that $s_\varepsilon\in L^1_{loc}({\mathbb R}^3)$ with 
$s_\varepsilon \rightarrow \frac{1}{r}\in{\mathcal D}'(\R^3)$. 
Moreover, we may require $o_\varepsilon(r)$ to be smooth on ${\mathbb R}^3\backslash \{0\}$.
Summing over (\ref{r00epsilon}) and (\ref{r22epsilon}) we
get $R_\varepsilon = 2m (\frac{1}{r}\, o_\varepsilon^{\prime\prime} +
\frac{2}{r^2}\, o_\varepsilon^\prime)\,$.
Choosing $o_\varepsilon(r) = (1 + c\, [r^\varepsilon-1])$ 
we obtain for $R_\varepsilon$ different weak limits as
the constant $c$ varies, i.e.,
\begin{equation}\label{arbitrarycurvature} 
R_\varepsilon \,\rightarrow\, 8 \pi m c \,\delta.
\end{equation}
For $o_\varepsilon = r^{-\varepsilon}$ the situation is even worse.
Although $o_\varepsilon \in L^1_{loc} \:\forall\, \varepsilon$ and 
$o_\varepsilon\rightarrow 1\in{\mathcal D}'$ as  $\varepsilon \rightarrow 0$,
$R_\varepsilon$ does not converge weakly, so that we obtain no 
distributional result whatsoever.

Nonetheless, similar non-smooth regularizations have been considered in the literature.
In these cases the desired result (\ref{ricciresults}) can only be produced by means
of a clever choice of explicit formulae; in particular,
$o_\varepsilon = r^\varepsilon$ in \cite{Pantoja/Rago:2000} and
$o_\varepsilon(r) = \Theta(r-\varepsilon)$ in \cite{Pantoja/Rago:1997}.
The authors of \cite{Balasin/Nachbagauer:1993} have shown that
the result (\ref{ricciresults}) may be reproduced as long as $o_\varepsilon|_{r=0} =0$.
However, the conceptual problem remains untouched: $R_\varepsilon[h_\varepsilon]$ can
only be derived for smooth regularizations $h_\varepsilon$; distributions cannot be used
as an input for nonlinear operations.
\vskip12pt

Prior to a more detailed investigation of the choice of 
regularization, we briefly comment on two more attempts 
in the literature.
In \cite{Pantoja/Rago:1997} a regularization of the metric
using thin shell solutions is investigated.
The limit $(\varepsilon \rightarrow 0)$ corresponds to
a shrinking of the shell. 
However, the shells can only be placed outside the horizon (of a 
black hole with identical mass).
This implies that a shrinking of the shell must be coupled to
a decrease in mass: $m$ converges to zero in the limiting process,
so the obtained results should either be considered trivial
($R \sim\, m \delta\,|_{m=0}= 0$) or be rejected completely.

In \cite{Kawai/Sakane:1997} the authors claim to have found
different results for the curvature quantities by
regularizing the Schwarzschild metric in a different coordinate
system. They study
the interrelations of regularizations and coordinate transformations
for this particular problem.
However, some details are not overly convincing.
If we choose a new radial coordinate $\tilde{r}$ such that
$r = \Lambda \tilde{r} + a$ with $a=2m$, then
$\tilde{r} = 0$ does not describe the Schwarzschild singularity.
Instead, $\tilde{r}=0$ corresponds to the
coordinate singularity at the horizon $r=2m$, but shrunk to
one point. Obviously, we should not compare the outcome
of these considerations with our former results.

\vskip12pt
We now begin with an in-depth analysis of certain aspects of the
regularization procedure commencing with the issue
of componentwise regularization and invertibility of the regularized metric.
According to (\ref{tensorp}) in section \ref{prereq}, regularizing a tensor such as the 
Schwarzschild metric (\ref{ssmetric}) comes up to regularizing each
distribution-valued component separately.
Following this rule we obtain a regularized metric slightly different from
(\ref{ssmetricembed}), namely
\begin{equation}\label{ssmetricembed2} 
ds^2_\varepsilon = 
h_\varepsilon(r)\: dt^2 - (h^{-1})_\varepsilon(r) \: dr^2 
+ r^2 \:d\Omega^2\,.
\end{equation}
Since $\cl[h_\varepsilon] \cl[(h^{-1})_\varepsilon]\not=1\in\gs$,
the determinant of the regularized metric~(\ref{ssmetricembed2})
is no longer identically one.
(This, in fact, does not come as a surprise; cf.\ Schwartz' impossibility result \cite{simp}.)
However, the product {\em is} preserved in the sense
of association, i.e., $h_\varepsilon(h^{-1})_\varepsilon\approx 1$.
Analogous issues arise from the inverse metric:
embedding also $g^{-1}$ componentwise into ${\mathcal G}$ we obtain regularized objects,
$g_\varepsilon$ and $(g^{-1})_\varepsilon$, which are only inverse to each other 
in the sense of association. Taking a different viewpoint, however, it is comparatively
easy to avoid these issues: on the classical level the Schwarzschild geometry is uniquely determined
by the set of variables $\{g_{tt}, g_{rr}, g_{\theta\theta}, g_{\varphi\varphi}\}$,
or, e.g., equivalently by $\{g_{tt}, g_{\theta\theta}, g_{\varphi\varphi}, \det g\}$.
Embedding the second set of variables into ${\mathcal G}$ leads directly to the
regularization (\ref{ssmetricembed}) used above;
no invertibility problems arise at all since $\det g_\eps$ is forced to equal one.
\vskip12pt

Finally we return to discussing the problem of smoothness
of the regularized metric from a different, more geometrical 
point of view. We regard this problem to be so 
essential that in the next section we propose an approach 
entirely different form the one taken so far.

In fact, the regularizations
used so far (as all the other regularizations in the relevant 
literature) do {\em not} 
provide a {\em smooth} regularized metric tensor. This fact is hidden again by the use
of spherical coordinates. In Cartesian coordinates pertaining to 
$(r,\theta,\phi)$---which we used to fix the differentiable structure
of the extended manifold at $r=0$---however, it can be 
explicitly seen from the form of the metric
\begin{equation}\label{ssmetriccartesform} 
ds^2 = h(r)\:dt^2 + d\vec{x}^2 -(1+h(r)^{-1}) \frac{x_i x_j}{r^2}\, dx^i dx^j \:.
\end{equation}
In order to obtain a smooth regularization it is not sufficient to merely regularize
$h(r)$. In fact, we must embed the singular coefficient
$(1+h(r)^{-1}) \frac{x_i x_j}{r^2}$ as a whole into ${\mathcal G}$.
Apart from technical difficulties this approach should provide a smooth
regularized metric $ds_\varepsilon$.
However, we have reached an impasse:
the regularized metric will not be invertible at some
distinct value $r_0$ of the radial coordinate, 
where $r_0 \rightarrow 0$ ($\varepsilon\rightarrow 0$).
This will be shown in the remainder of this section.

As shown in appendix \ref{appendixB}, the regularization of (\ref{ssmetriccartesform}) 
takes the form 
\begin{equation}\label{ssmetriccartesformembed2} 
ds^2_\varepsilon = h_\varepsilon(r)\:dt^2 + (1-a_\varepsilon(r))\: dr^2 + 
(1-b_\varepsilon(r))\: r^2 d\Omega^2,
\end{equation}
with $a_\varepsilon(0) \rightarrow \frac{1}{3}$ ($\varepsilon\rightarrow 0$).
This implies that the $rr$-component of the regularized 
metric~(\ref{ssmetriccartesformembed2}) is positive at $r=0$ (at least for small
$\varepsilon$), because 
$(g_{rr})_\varepsilon(0) = (1-a_\varepsilon(0)) \rightarrow \frac{2}{3}$ 
($\varepsilon\rightarrow 0$).
On the other hand, $(1-a_\varepsilon)$ approximates $-h^{-1}$, i.e.,
$(g_{rr})_\varepsilon(r\neq 0) \rightarrow -\frac{r}{2m-r} < 0 \quad 
(\varepsilon \rightarrow 0).$
So we conclude that at some value $r_0$ of $r$
the smooth function $(g_{rr})_\varepsilon(r)$ must have a zero
at least for small $\varepsilon$. (Interestingly enough, this is not the 
case for negative masses since $-\frac{r}{2m-r}$ is positive then). 
In other words, this means that the regularization of the metric~(\ref{ssmetriccartesform})
degenerates at some radius $r_0$.
Evidently, $r_0 \rightarrow 0$ as $\varepsilon\rightarrow 0$.

Note that the occurrence of this radius of degeneracy is neither due to the fact 
that we choose the particular regularization~(\ref{ssmetriccartesformembed2}), nor is it 
possible to avoid it by giving up spherical symmetry. To see this in some more detail 
consider the spatial part of (\ref{ssmetriccartesform})
(set $\tilde{h}(r):= h^{-1}(r)$) and consider a certain class of regularizations
\begin{equation}\label{reg1} 
ds^2_\varepsilon = d\vec{x}^2 - (1+\tilde{h}_\varepsilon(r))\,
 \frac{x_i x_j}{r^2}\, dx^i dx^j,
\end{equation}
where $\tilde{h}_\varepsilon(r)$ denotes an arbitrary regularization of $\tilde{h}(r)$.
However, for $ds^2_\varepsilon$ to become smooth, we must require that $\tilde{h}_\varepsilon(r)$ 
be $-1 + O(r^2)$ for $(r\rightarrow 0)$. Now, an arbitrary regularization of $ds^2$ not 
necessarily respecting spherical symmetry is obtained by adding zero-sequences to~(\ref{reg1}). 
\begin{equation}\label{notsphersymmreg} 
ds^2_\varepsilon = d\vec{x}^2 - (1+\tilde{h}_\varepsilon(r))\,
\frac{x_i x_j}{r^2}\, dx^i dx^j + (a_{ij})_\varepsilon(\vec{x})\,dx^i dx^j.
\end{equation}
For special cases of $(a_{ij})_\varepsilon$ it is easy to show that (\ref{notsphersymmreg}) 
is degenerate. Choose, e.g., $(a_{ij})_\varepsilon$ such that only $(a_{12})_\varepsilon 
=:b_\varepsilon(\vec{x})$ is non-vanishing. The determinant of (\ref{notsphersymmreg}) at 
$\vec{x} = (0,0,r)$
is equal to $-\tilde{h}_\varepsilon(r)(1-b_\varepsilon(\vec{x})^2)$. 
As $-\tilde{h}_\varepsilon(0) = 1$ and $-\tilde{h}_\varepsilon(r) < 0$
for small $\varepsilon$ and finite $r$, there exists a radius $r_\varepsilon$ 
(with $r_\varepsilon \rightarrow 0$
for $\varepsilon \rightarrow 0$), such that $\det(g_\varepsilon) = 0$.
Again, we observe degeneracy.
\vskip12pt

\section{The Kerr-Schild approach}\label{thekerrschildapproach} 

To begin with let us summarize what we have done so far:
we considered regularizations of the Schwarzschild metric (using the language of
algebras of generalized functions) to calculate the (distributional) curvature
at the singularity. 
The regularizations used were essentially based on Cartesian coordinates
associated with the spherical Schwarzschild coordinates.
However, it turned out that all these regularizations were either
non-smooth or not invertible.
{\em Smoothness and invertibility mutually exclude each other} in this context. 
Hence, we are going to take another more
geometrical view-point in this section. 
The main idea---following \cite{parker,Balasin/Nachbagauer:1994}---is to use the Kerr-Schild 
form of the Schwarzschild metric. Retaining this preferred structure also 
during the whole regularization process will enable us to derive the physically desired
result in a rigorous manner.

A metric belongs to the so-called Kerr-Schild class \cite{Debney/Kerr/Schild:1969}
if it can be written as 
\begin{equation} 
g_{ij} = \eta_{ij} + f k_i k_j \qquad \mbox{with} \quad k^i k_i = 0.
\end{equation}
Here, the null vector field $k\in{\mathfrak X}$ is normalized ($k^0 = 1$) and $f$
is a smooth function.
Exploiting the Kerr-Schild form, some curvature quantities take a 
particularly simple form, e.g.,
\begin{equation}\label{ksscalar} 
R=\partial_a\partial_b(f k^a k^b).
\end{equation}
The Schwarzschild metric is a member of the Kerr-Schild class.
In fact, transformation to Eddington-Finkelstein coordinates
($\bar{t} = t + 2m \log |2m-r|$) yields
\begin{equation}\label{kss} 
ds^2 = -d\bar{t}^2 + dr^2 + r^2 d\Omega^2 + \frac{2m}{r} (d\bar{t}-dr)^2.
\end{equation}
Evidently, (\ref{kss}) is of Kerr-Schild form, $\,g_{ij} = \eta_{ij} + f k_i k_j\,$, with
\begin{equation}\label{ksform} 
k = \frac{\partial}{\partial \bar{t}} + \frac{\partial}{\partial r}
\quad\mbox{ and }\quad
f = \frac{2m}{r}\,.
\end{equation}
Analogously to section~\ref{afirstapproachtotheproblem}, we regard $f$ and $k$
as distributions on $\R^4$. By this we again implicitly fix the differentiable structure: 
the coordinates $(\bar{t}, x^i)$ are extended through the origin. 

We now proceed by regularizing {\em both} $f$ and $k$. Indeed, this is necessary due to the
fact that not only the profile function $f$ is singular, but also the null vector field $k$ is
non-smooth. Recall that, on account of the nonlinearities in $R[g]$,
(\ref{ksscalar}) can only be derived for smooth functions; it is inaccessible for
distributional input. (Note the analogy of this situation with the one encountered in
(\ref{ssmetricembed})). Hence, $f$ and $k$ are chosen to be the fundamental 
variables characterizing the metric (compare with the remarks in Section \ref{com+prob}).
Regularizing the function $f$ as in Section \ref{afirstapproachtotheproblem} gives
\begin{equation}\label{fandkembed} 
f(r)=\dfrac{2m}{r}\stackrel{\iota}{\rightarrow}\iota(f(r)) = 
f_\varepsilon(r)= 2m (\dfrac{1}{r})_\varepsilon\,.
\end{equation}
The regularization of $k_i$ is carried out in detail in appendix \ref{appendixC},
yielding
\begin{equation}\label{ntc}
k_i(r) = \dfrac{x_i}{r} \stackrel{\iota}{\longrightarrow}
\iota(\dfrac{x_i}{r}) = k_{i\,\varepsilon} = 
(\dfrac{x_i}{r})_\varepsilon=x_i \,F_\varepsilon(r)
\qquad (i=1,\ldots,3)\,,
\end{equation}
where $F_\varepsilon$ is given by (\ref{C3}).
Note that, for the moment, $k_0 = 1$ is embedded trivially into ${\mathcal G}$.
Collecting the results of (\ref{fandkembed}) and (\ref{ntc}) we get the regularized
metric
\begin{equation}\label{embedks!} 
ds_\varepsilon^2 = (-1 + f_\varepsilon) dt^2 - 2 r f_\varepsilon F_\varepsilon\, dt dr +
(1+ \tilde{f}_\varepsilon) dr^2 + r^2 d\Omega^2\:,
\end{equation}
where, for simplicity, $\bar{t}$ has been replaced by $t$ again, and
$\tilde{f}_\varepsilon$ abbreviates $r^2 f_\varepsilon F_\varepsilon^2$.
Unfortunately, (\ref{embedks!}) is no longer of Kerr-Schild form.
This can be shown indirectly:
assuming that (\ref{embedks!}) is Kerr-Schild, i.e., assuming that 
$ds_\varepsilon^2 = -dt^{\prime\,2} + dr^2 + r^2 d\Omega^2 + 
f_\varepsilon^\prime\, (dt^\prime - dr)^2\,$ can be achieved by
a transformation $t\rightarrow t^\prime(t,r)\,$, 
it follows that $
f_\varepsilon^\prime = 
\frac{\tilde{f}_\varepsilon}{1- f_\varepsilon + \tilde{f}_\varepsilon}$ and
$(\frac{dt}{dt^\prime})^2 = 
\frac{1}{1-f_\varepsilon+ \tilde{f}_\varepsilon}$.
As a consequence, the denominator $1-f_\varepsilon+ \tilde{f}_\varepsilon$
must be a strictly positive function. In fact, in the sense of association, it is
equal to one. However, 
at the origin $r=0$, we obtain $1-f_\varepsilon+\tilde{f}_\varepsilon = 
1- 8\pi m \frac{1}{\varepsilon} \int_0^\infty t \rho(t) dt\:$, which
is negative for small $\varepsilon$ as long as $\int_0^\infty t \rho(t) dt> 0$,
a contradiction.

The fact that the embedding (\ref{embedks!}) is no longer of Kerr-Schild form
bears a strong relation to the fact that smooth regularizations degenerate
at a certain value of $r$ (see Section \ref{com+prob}): the determinant of~(\ref{embedks!}) 
contains the factor $1-f_\varepsilon+\tilde{f}_\varepsilon\,$, which
was shown to possess a zero.

Additionally, in analogy to the statements made at the end of Section \ref{com+prob},
we note that the loss of the Kerr-Schild form does not stem from choosing to 
regularize the singular coefficient functions via convolution. On the 
contrary, it may be shown that any regularization of the metric displays this behavior as long as 
only the spatial components of $k$ are taken into account.

We will now take the announced geometrical view-point:
we consider regularizations retaining the Kerr-Schild decomposition.
This requires, in particular, that the regularized vector $k_\varepsilon$ is still null.
Thus, we consider the regularization
\begin{equation}\label{allkcomponentsembed} 
k_\varepsilon^i = r F_\varepsilon\, k^i \qquad (i=0\ldots 3)\,.
\end{equation} 
While the spatial components of~(\ref{allkcomponentsembed}) coincide 
with~(\ref{ntc}), $k^0_\varepsilon$ is only associated to 1
($k_\varepsilon^0 = r F_\varepsilon\approx 1$).
As required, $k_\varepsilon$ satisfies the condition $k_\varepsilon^i k_{\varepsilon\,i}=0$.
Note that, in order to obtain~(\ref{allkcomponentsembed}),
the functions $f$, $k^i$ ($i=1\ldots 3$) and $k\cdot k$ are chosen as
fundamental variables determining the geometric structure of the spacetime.

Using (\ref{fandkembed}) and (\ref{allkcomponentsembed}) the regularized metric takes the form
\begin{equation}\label{ksembed2}
g_{ij\,\varepsilon} = 
\eta_{ij} + f_\varepsilon k_{\varepsilon\,i} k_{\varepsilon\,j} = 
\eta_{ij} + (r^2 F_\varepsilon^2 f_\varepsilon) \,k_i k_j =
\eta_{ij} + \tilde{f}_\varepsilon \, k_i k_j.
\end{equation}
Obviously, (\ref{ksembed2}) is of Kerr-Schild form.
Finally we have arrived at a regularization of the
Schwarzschild metric which is {\em both smooth and invertible}
(the inverse being $\eta^{ij} - \tilde{f}_\varepsilon k^i k^j$).
This allows us to fully exploit the Kerr-Schild form, i.e.,
to use (\ref{ksscalar}), to obtain
\begin{equation}\label{kscurvature} 
R_\varepsilon = \partial_a\partial_b(\tilde{f}_\varepsilon k^a k^b) =
\frac{4}{r}\, \tilde{f}_\varepsilon^\prime + 
\tilde{f}_\varepsilon^{\prime\prime} + \frac{2}{r^2}\,\tilde{f}_\varepsilon\,.
\end{equation}
To complete our program we calculate the weak limit of $R_\eps$. The technically
involved calculations are deferred to appendix \ref{appendixD}. Finally we derive
\begin{equation}\label{finalresult} 
(R_\varepsilon)_\varepsilon \:\approx\: 8 \pi m \delta\,.
\end{equation}
The Ricci tensor can be treated in complete analogy
to obtain the Einstein tensor
\begin{equation}
(G^a_{\:b\:\varepsilon})_\varepsilon \;\approx\; -8 \pi m\,\delta\:\delta^a_0\delta^0_b \:.
\end{equation}

\section{The horizon}\label{thehorizon} 

In this last section we leave the neighborhood of the singularity
at the origin and turn to the singularity at the horizon.
The question we are aiming at is the following:
using distributional geometry (thus without leaving Schwarzschild coordinates),
is it possible to show that the horizon singularity of the Schwarzschild metric
is merely a coordinate singularity?
In order to investigate this issue we calculate the distributional curvature at the horizon
(in Schwarzschild coordinates).

Examining the Schwarzschild metric~(\ref{ssmetric}) in
a neighborhood of the horizon,
we see that, whereas $h(r)$ is smooth, $h^{-1}(r)$ is not even $L^1_{loc}$
(note that the origin is now always excluded from our considerations; the space we 
are working on is $\mathbb{R}^3\backslash\{0\}$).
Thus, regularizing the Schwarzschild metric amounts to embedding $h^{-1}$ into
${\mathcal G}$ (as done in~(\ref{vpembed})).
\begin{equation}\label{horizonmetric}
ds^2_\varepsilon = 
h(r)\: dt^2 - (h^{-1})_\varepsilon(r)\: dr^2 + r^2 \:d\Omega^2
\end{equation}
\begin{equation}
\mbox{with}\quad
h(r) = -1 +\frac{2m}{r} \quad\;\mbox{and}\quad\;
(h^{-1})_\varepsilon(r)= -1-2m [\vp(\frac{1}{r-2m})]_\varepsilon
\end{equation}
Obviously, (\ref{horizonmetric}) is degenerate at $r=2m$, because 
$h(r)$ is zero at the horizon.
However, this does not come as a surprise.
Both $h(r)$ and $h^{-1}(r)$
are positive outside of the black hole and 
negative in the interior.
As a consequence {\em any}
(smooth) regularization of $h(r)$ (or $h^{-1}$) 
must pass through zero somewhere and, additionally, this zero 
must converge to $r=2m$ as the regularization parameter goes to zero
(note the analogy to the situation in section~\ref{com+prob}).

Due to the degeneracy of~(\ref{horizonmetric}),
the Levi-Civit\`a connection is not available.
Consider, therefore, the following connection $\Gamma^l_{kj} \in {\mathcal G}$:
\begin{equation}\label{horconn}
\Gamma^l_{kj} =\frac{1}{2} [\iota(g^{-1})]^{\,lm}
[\iota(g)_{mk,j}+\iota(g)_{mj,k}-\iota(g)_{kj,m}]
\end{equation}
Clearly, $\Gamma$ coincides with the Levi-Civit\`a connection
on ${\mathbb R}^3\backslash\{r=2m,r=0\}$, as
$\iota(g)=g$ and $\iota(g^{-1})=g^{-1}$ there.

Unfortunately, $\Gamma$ does not respect the regularized 
metric $\iota(g)$
(\ref{horizonmetric}), i.e.,
$\iota(g)_{ij;k} \neq 0$, e.g.,
$\iota(g)_{00;1} = (1-h (h^{-1})_\varepsilon) h^\prime$.
However, 
compatibility with the metric $\iota(g)$ is 
a priori ruled out by the following statement:
there exists no connection whatsoever under which
$\iota(g)$ would be a parallel tensor.
To show this, just look at ($L^i_{\:jk}$ denoting a not necessarily
torsion-free connection)
$\iota(g)_{00;1} = \iota(g)_{00,1} - 2 L^0_{\:10} \iota(g)_{00}\,$.
At the horizon $\iota(g)_{00} = 0$, so that 
$\iota(g)_{00;1}|_{r=2m} = h^\prime(2m) = -\frac{1}{2m} \neq 0$.
In the sense of association, however, the connection (\ref{horconn})
is in fact metric compatible:
$\iota(g)_{ij;k} \approx 0$.

We now investigate 
the curvature pertaining to the connection (\ref{horconn}), 
picking out $R_{00\,\varepsilon}$ as a characteristic example.
The result of the calculations displays the following structure

\begin{eqnarray}\label{horricci0}
R_{00\,\varepsilon}(r) &=&
\vp_\varepsilon^\prime (-\frac{m^2}{r^2}+4\frac{m^3}{r^3} - 4\frac{m^4}{r^4})+ 
\vp_\varepsilon (2\frac{m^3}{r^4}- 4\frac{m^4}{r^5}) + 
(-\frac{m^2}{r^4}-2\frac{m^3}{r^5})\\
&=& 
\vp_\varepsilon^\prime(r)\sum\limits_{l=2}^\infty c_l x^l + 
\vp_\varepsilon(r)\sum\limits_{l=1}^\infty d_l x^l 
- \frac{1}{8m^2} - \frac{1}{16m^2} \sum\limits_{l=1}^\infty e_l x^l
\label{horricci}
\end{eqnarray}
Here, the abbreviations 
$\vp_\varepsilon =[\vp(\frac{1}{r-2m})]_\varepsilon$ and $x=\frac{r-2m}{2m}$
are used; $c_l$, $d_l$ and $e_l$ are constants.
Equation (\ref{horricci}) holds for $|x|<1$; 
the infinite sums converge in this case.

If the horizon is excluded, $R_{00\,\varepsilon} = 0$ (mod ${\mathcal N}$),
because (\ref{horconn}) coincides with the 
Schwarzschild Levi-Civit\`a connection there. 
In the neighborhood of $r=2m$ we aim at comparing $R_{00\,\varepsilon}(r)$
with a Colombeau object of the type 
$f(\frac{r-2m}{\varepsilon})$
($f$ a Schwartz function). 
To this end we choose a fundamental sequence
$r_\varepsilon = 2m + \varepsilon^q a_0$ converging to $r=2m$ 
and examine $R_{00\,\varepsilon}(r_\varepsilon)\,$
(use~(\ref{horricci}) together with~(\ref{vpembed})).
\begin{itemize}
\item $q>1$: 
$R_{00\,\varepsilon}(r_\varepsilon) = \mbox{const} + o(\varepsilon^{q-1})$. 
\item $q<1$: Using (\ref{vpembed}) we find that
$\vp_\varepsilon(r_\varepsilon) = \frac{1}{r_\varepsilon-2m}$
and $\vp_\varepsilon^\prime(r_\varepsilon) = -\frac{1}{(r_\varepsilon-2m)^2}$
(in the sense of generalized numbers). 
Inserting these results into equation (\ref{horricci0}), we
obtain $R_{00\,\varepsilon}(r_\varepsilon) = 0$.
\item $q\:=\:1$: $R_{00\,\varepsilon}(r_\varepsilon) = \mbox{const} + o(1)$.
\end{itemize}

Thus, 
$R_{00\,\varepsilon}(r_\varepsilon)$ has the same asymptotic behavior as
a sequence of the type 
$f(\frac{r_\varepsilon-2m}{\varepsilon})$ (as $\varepsilon \rightarrow 0$).
As a consequence, the weak limit of  $R_{00\varepsilon}(r)$ 
can be calculated easily, simply by evaluating  
$\int dr\, r^2 \tilde{\Phi}(r) f(\frac{r-2m}{\varepsilon})$.
Evidently, this expression vanishes as $\epsilon \rightarrow 0$.
Since analogous results hold for the other components of the 
Ricci tensor,
we are finally able to state
\begin{equation}\label{riccihorizon0} 
R_{ij\,\varepsilon} \approx 0\,.
\end{equation}

In other words: viewed as a distribution, $R_{ij} = 0$ on 
${\mathbb R}^3\backslash\{0\}$, i.e., including the
horizon.
If we were courageous enough we could take this as a proof that
the metric singularity at the horizon is only a 
coordinate singularity.

We conclude this section with a remark on the connection~(\ref{horconn}).
Due to the degeneracy of any regularization of the metric (e.g. (\ref{horizonmetric})) no
canonical (Levi-Civit\`a) connection could be defined.
The choice of connection~(\ref{horconn}) bears a strong relation to the
regularized metric; however, there seems no way of telling if this choice
is canonical in some sense and thus preferred to other choices.
Despite this open question, 
at least it is clear that
the connection (\ref{horconn}) is a
regularization of the Schwarzschild {\em connection}.
Indeed, we could change our viewpoint: 
we consider the Schwarzschild connection (forgetting where it came from, i.e.,
forgetting about the metric),
regularize its distribution-valued components and calculate the distributional
curvature from it.
Proceeding in this manner, we obtain the result~(\ref{riccihorizon0}),
i.e., the spacetime is weakly Ricci-flat
(the origin was excluded from our considerations).

\begin{appendix}%
\section{Mollifier integrals}\label{appendixA:mollifierintegrals} 

Throughout this paper we work invariably with radially symmetric
mollifiers $\rho(r)$ (cf.\ section \ref{prereq}).
Most importantly, we have the properties
\begin{equation}\label{mollifierid} 
\int\limits_0^\infty dt\,t^2 \rho(t) = \frac{1}{4\pi}  \qquad\qquad 
\int\limits_0^\infty dt\,t^{2k} \rho(t) = 0 \:\: (k>1)\,.
\end{equation}

We investigate multiple integrals involving the mollifier
$\rho(r)$ and powers of $r\,$:
\begin{equation}\label{mollifierintegral1}
\int_x^\infty dt\, t^n \int_t^\infty s \rho(s)\, ds = 
-\frac{x^{n+1}}{n+1} \int_x^\infty t \rho(t) dt + 
\frac{1}{n+1} \int_x^\infty dt\, t^{n+2} \rho(t) 
\end{equation}

(\ref{mollifierintegral1}) holds for ($n,k \neq -1$), it is proven by simply performing 
integration by parts. 

One of the most interesting cases resulting from (\ref{mollifierintegral1})
is $n=0$ and $x\rightarrow 0\,$:
\begin{equation}\label{mollifierintegral3}
\int_0^\infty dt \int_t^\infty s \rho(s) ds = \frac{1}{4\pi}
\end{equation}

\section{Embedding of the Cartesian components}\label{appendixB} 
Referring to section \ref{com+prob} 
we investigate  
\begin{equation}\label{completeembedding} 
\iota\left( \frac{1+h(r)^{-1}}{r^2} x_i x_j\right) dx^i dx^j\:=\:
\left( 2m \int\limits \mbox{f}(\|\vec{z}\|)\, z_i z_j \,
\rho_\varepsilon(\|\vec{z}+\vec{x}\|)\, d^3z \right) dx^i dx^j\,,
\end{equation}
where $f(q) = \frac{1}{2m-q} \,\frac{1}{q^2}$.
In order to simplify (\ref{completeembedding}) we show the following relation
\begin{equation}\label{iotasmooth}
\iota\left(\frac{1+h(r)^{-1}}{r^2} x_i x_j\right) =
x_i\, x_j\:c_\varepsilon(\vec{x}) \quad\: \mbox{for}\:\, i\neq j 
\qquad\quad (c_\varepsilon \mbox{ smooth}).
\end{equation}
Proof: Since both $f(\|\vec{z}\|)$ and $\rho_\varepsilon$ are
even functions in $z_i\,$, we observe that
\begin{equation}
\iota(\tfrac{1+h(r)^{-1}}{r^2} x_i x_j)\,\Big|_{x_i=0} = 
2m \int
f(\|\vec{z}\|) z_i z_j \rho_\varepsilon(\ldots,z_i,\ldots) d^3z
= 0\,.
\end{equation}
We can conclude that 
\begin{equation}
\begingroup\textstyle
\iota(\frac{1+h(r)^{-1}}{r^2} x_i x_j) = 
x_i \,c^\prime_\varepsilon(\vec{x})
\endgroup \qquad\quad (c^\prime_\varepsilon \mbox{ smooth})
\end{equation}
Also, $\iota(\frac{1+h(r)^{-1}}{r^2} x_i x_j)|_{x_j=0} = 0\:$, from
which follows that 
$c^\prime_\varepsilon(\vec{x}) = x_j c_\varepsilon(\vec{x})$,
yielding (\ref{iotasmooth}). 
Note, however, that the smooth function $c_\varepsilon(\vec{x})$ 
in (\ref{iotasmooth}) is
not equal to $\iota(\tfrac{1+h(r)^{-1}}{r^2})$. \proofend

In the case $i=j$, equation (\ref{iotasmooth})
is no longer valid. We are able to calculate
the $ii-$component in the limiting case $\varepsilon\rightarrow 0$, i.e.,
\begin{equation}\label{iicomp}
\iota\left(\frac{1+h(r)^{-1}}{r^2} x_i^2\right)\Big|_{\vec{x}=0} = 
2m \int
f(\|\vec{z}\|) z_i^2 \rho_\varepsilon(\|\vec{z}\|) d^3z
\;\rightarrow\; \frac{1}{3} \quad\;(\varepsilon\rightarrow 0).
\end{equation}
Proof: Clearly, $\,2m \int
f(\|\vec{z}\|) z_i^2 \rho_\varepsilon(\|\vec{z}\|) d^3z$ 
is independent of the choice of the index $i$, so that
we may substitute it by 
$\frac{2m}{3}\int
f(\|\vec{z}\|) \|\vec{z}\|^2 \rho_\varepsilon(\|\vec{z}\|) d^3z$.
Obviously, this converges to $\frac{1}{3}$ as $\varepsilon$ goes to zero.
\proofend

The regularized metric (\ref{completeembedding}) 
is radially symmetric, $R^* g_\eps = g_\eps$ 
($R$ a rotation), as long as radially symmetric mollifiers are used.
Thus, it must be of the form of a general
radially symmetric metric 
$ds^2 = a(r)\: dr^2 + r^2 b(r)\:d\Omega^2$, hence 
\begin{equation}\label{generalradsymm}
\iota\left(\frac{1+h(r)^{-1}}{r^2} x_i x_j\right) dx^i dx^j =
(a_\varepsilon-b_\varepsilon)(r)
\frac{x_i}{r}\frac{x_j}{r} dx^i dx^j + b_\varepsilon(r)\:d\vec{x}^2\,.
\end{equation}

For the general radially symmetric metric (\ref{generalradsymm}) to be smooth 
$(a_\varepsilon-b_\varepsilon)(r) = O(r^2)$ is required.
We observe consistency with (\ref{iotasmooth}) and conclude
\begin{equation}
a_\varepsilon(r) = b_\varepsilon(r) + c_\varepsilon r^2\,.
\end{equation}
At the origin $r=0$ only the second term 
$b(r)\:d\vec{x}^2$ remains relevant, since 
$(a-b)(r)\frac{x_i}{r}\frac{x_j}{r}|_{r=0} = 
c_\varepsilon x_i x_j |_{r=0} = 0$.
Turning equation (\ref{iicomp}) to good account we obtain
\begin{equation}\label{aonethird}
b_\varepsilon(0) \rightarrow \frac{1}{3} \quad (\varepsilon\rightarrow 0), \qquad
a_\varepsilon(0) \rightarrow \frac{1}{3} \quad (\varepsilon\rightarrow 0)\,.
\end{equation} 

Combining (\ref{ssmetriccartesform}) with (\ref{generalradsymm}), we finally obtain
\begin{equation}
ds^2_\varepsilon = h_\varepsilon(r)\:dt^2 + (1-a_\varepsilon(r)) dr^2 + 
(1-b_\varepsilon(r)) r^2 d\Omega^2 \,.
\end{equation}

\section{Embedding of $k^i$}\label{appendixC} 
We explicitly embed the radially outward pointing unit vector field 
$k_i = \frac{x_i}{r}$ ($i=1\cdots 3$)
into the Colombeau algebra, i.e.,
\begin{equation}\label{kembed} 
\iota(\frac{x_i}{r}) = 
\int d^3x^\prime\frac{x_i-x_i^\prime}{\|\vec{x}-\vec{x}^\prime\|}\,
\rho_\varepsilon(\|\vec{x}^\prime\|) = 
-\int d^3z\,\frac{z_i}{\|\vec{z}\|} \,
\rho_\varepsilon(\|\vec{x}+\vec{z}\|)\,.
\end{equation}

Equation (\ref{kembed}) is of an analogous form 
as (\ref{completeembedding})
in appendix \ref{appendixB}. We may conclude 
that $\iota(\frac{x_i}{r})$ is a radially
symmetric vector field. Moreover,
despite $\iota(\frac{x_i}{r}) \neq x_i \iota(\frac{1}{r})\,$,
we must still have  
(repeating (\ref{iotasmooth})ff.) 
\begin{equation}\label{xioverrembed} 
\iota(\frac{x_i}{r}) = x_i\, F_\varepsilon(\vec{x})\,\qquad (i=1\ldots 3)\:.
\end{equation}
Here, $F_\varepsilon(\vec{x})$ is a smooth function and moreover, 
because this function 
must be radially symmetric, $F_\varepsilon(\vec{x}) = F_\varepsilon(r)$.

This fact makes it possible to calculate $\iota(\frac{x_i}{r})$ 
explicitly. Take $x=(0,0,r)$ and investigate
$\iota(\frac{x_3}{r}) = x_3 F_\varepsilon(r) = r F_\varepsilon(r)\,$:
\begin{eqnarray}\label{C3}\nonumber
r F_\varepsilon(r) & = & 
\int \frac{x_3-x_3^\prime}{\|\vec{x}-\vec{x}^\prime\|} 
\,\rho_\varepsilon(r^\prime) \,d^3x^\prime \\\nonumber
& = & 
2\pi \int r^{\prime\,2} dr^\prime \,\rho_\varepsilon(r^\prime)
\int_{-1}^1 d(\cos\theta^\prime) \,
\frac{r-r^\prime \cos\theta^\prime}{\sqrt{r^2+r^{\prime\,2}-
2 r r^\prime\cos\theta^\prime}} \\\nonumber
& = & 
r (\frac{1}{r})_\varepsilon - 2\pi  \int r^{\prime\,2} dr^\prime \,\rho_\varepsilon
\left( \frac{1}{r} (|r-r^\prime| + (r+r^\prime)) +
\frac{1}{3}\frac{1}{r^2}\frac{1}{r^\prime} (|r-r^\prime|^3 - (r+r^\prime)^3)
\right) \\
\label{Fepsilon} 
& = & 
r \left(  
\frac{4\pi}{r} \int\limits_0^r ds\, s^2 \rho_\varepsilon(s) +
\frac{8 \pi}{3} \int\limits_r^\infty ds\, s \rho_\varepsilon(s) -
\frac{4\pi}{3}\,\frac{1}{r^3} \int\limits_0^r ds\, s^4 \rho_\varepsilon(s)
\right)
\end{eqnarray}  

Clearly, $F_\varepsilon(r) = \frac{1}{r}$ on ${\mathbb R}^3\backslash \{0\}$
and, moreover, $F_\varepsilon(r) \approx \frac{1}{r}$ on the whole space.
We can write the latter also in the form $r F_\varepsilon(r) \approx 1$.

\section{Weak limits for the Kerr-Schild case}\label{appendixD}

We investigate the distributional limit of (\ref{kscurvature}).
Inserting for $\tilde{f}_\varepsilon$, (\ref{kscurvature}) becomes
\begin{eqnarray}\label{almostcurvscal2explicit} \nonumber
R_\varepsilon & =  &
\frac{128 \pi^3 m}{3} \:\Big(\:
16 [1]_\varepsilon^3 + \frac{4}{r^7} [2]_\varepsilon [4]_\varepsilon^2 + 
\frac{2}{r^6} [1]_\varepsilon [4]_\varepsilon^2 - 
\frac{4}{r^5} [4]_\varepsilon [2]_\varepsilon^2 + \\ \nonumber
& & \qquad\quad +
\frac{32}{r} [2]_\varepsilon [1]_\varepsilon^2 + 
\frac{14}{r^2} [1]_\varepsilon [2]_\varepsilon^2 - 
3 \rho_\varepsilon [2]_\varepsilon^2 - 
4 r \rho_\varepsilon [1]_\varepsilon [2]_\varepsilon +\\ 
& & \qquad\quad  -
\frac{4 r^2}{3} \rho_\varepsilon [1]_\varepsilon^2 - 
\frac{1}{3 r^4} \rho_\varepsilon [4]_\varepsilon^2 +
2 \frac{1}{r^2} \rho_\varepsilon [2]_\varepsilon [4]_\varepsilon + 
\frac{4}{3r} \rho_\varepsilon [1]_\varepsilon [4]_\varepsilon
\:\Big)\:,
\end{eqnarray}
where $[1]_\varepsilon:=\int\limits_r^\infty ds \,s \rho_\varepsilon(s)$,
$[2]_\varepsilon:=\int\limits_0^r ds\,s^2 \rho_\varepsilon(s)$,
$[4]_\varepsilon:=\int\limits_0^r ds\,s^4 \rho_\varepsilon(s)$.

In order to compute the weak limit of~(\ref{almostcurvscal2explicit}),
expressions of the form (\ref{ijk2}) and (\ref{ijk3}) below have to be investigated.
(Note that the negative powers of $r$ are compensated
by the integrals so that 
both (\ref{ijk2}) and (\ref{ijk3}) are well-defined
as $r\rightarrow 0$).
\begin{equation}\label{ijk2} 
r^{-4+2j+3i}\: \rho_\varepsilon(r)\:\:
[1]_\varepsilon^{\:i}\: [2]_\varepsilon^{\:j} \: [4]_\varepsilon^{\:k} 
\qquad ( i+j+k = 2 )
\end{equation}
\begin{equation}\label{ijk3}
r^{-9+2j+3i}\:\: [1]_\varepsilon^{\:i}\: [2]_\varepsilon^{\:j} \: [4]_\varepsilon^{\:k} 
\qquad ( i+j+k = 3 )
\end{equation}
Terms of the forms (\ref{ijk2}) and (\ref{ijk3})
possess related distributional limits: 
\begin{eqnarray*}
\left(r^{-9+2j+3i}\,
[1]_\varepsilon^{\:i}\, [2]_\varepsilon^{\:j} \, [4]_\varepsilon^{\:k} 
\right)_\varepsilon 
& \approx  &
+ \frac{i}{2j+3i-6} 
\left(r^{-7+2j+3i} \rho_\varepsilon(r) 
[1]_\varepsilon^{\:i-1}\, [2]_\varepsilon^{\:j} \, [4]_\varepsilon^{\:k}
\right)_\varepsilon + \\
& & - \frac{j}{2j+3i-6}  
\left(
r^{-6+2j+3i} \rho_\varepsilon(r) 
[1]_\varepsilon^{\:i}\, [2]_\varepsilon^{\:j-1} \, [4]_\varepsilon^{\:k}
\right)_\varepsilon + \\
& & - \frac{k}{2j+3i-6}
\left( 
r^{-4+2j+3i} \rho_\varepsilon(r) 
[1]_\varepsilon^{\:i}\, [2]_\varepsilon^{\:j} \, [4]_\varepsilon^{\:k-1}
\right)_\varepsilon\,.
\end{eqnarray*}
In order to show this, consider the distributional action
on a test function $\Phi(\vec{x})$, i.e.,  
$\int\limits_0^\infty dr \,\tilde{\Phi}(r) r^{-7+2j+3i}
\, [1]_\varepsilon^{\:i}\, [2]_\varepsilon^{\:j} \, [4]_\varepsilon^{\:k}$. 
Here, we have introduced   
$\tilde{\Phi}(r) := \int \sin\theta d\theta d\phi \:\Phi(\vec{x})$.
Integrating by parts and using $\tilde{\Phi}^\prime(0) = 0$, 
the claim is established.

Taking $(j=2; k=1)$ as an example, we obtain the following weak limit:
\begin{eqnarray*}
\int\limits dr \,\tilde{\Phi}(r) \frac{1}{r^3}\,
[2]_\varepsilon^{\:2} [4]_\varepsilon 
&\:\stackrel{\varepsilon\rightarrow 0}{\:\longrightarrow\:}\:&
\tilde{\Phi}(0) \int\limits dx \rho(x) [2] [4] +
\frac{1}{2} \tilde{\Phi}(0)  \int\limits dx \,x^2 \rho(x) [2]^2 \\
& \stackrel{(\ref{mollifierid})}{=} & 
\tilde{\Phi}(0) \int\limits dx \rho(x) [2] [4] +
\frac{1}{6} \frac{1}{64 \pi^3} \tilde{\Phi}(0) \:.
\end{eqnarray*}
\begin{tabbing}
Here, \qquad  \=
$[1]_\varepsilon:=\int\limits_r^\infty ds \,s \rho_\varepsilon(s)$ \quad\=
$[2]_\varepsilon:=\int\limits_0^r ds\,s^2 \rho_\varepsilon(s)$ \quad\=
$[4]_\varepsilon:=\int\limits_0^r ds\,s^4 \rho_\varepsilon(s)$ \quad\kill  
Here,  \> 
$[1] :=\int\limits_x^\infty dt \,t \rho(t)$ \>
$[2] :=\int\limits_0^x dt\,t^2 \rho(t)$ \>
$[4] :=\int\limits_0^x dt\,t^4 \rho(t)\:$. 
\end{tabbing} 
Eventually, we obtain the result (\ref{finalresult}) for the
distributional limit of (\ref{almostcurvscal2explicit}).
\end{appendix}

\subsection*{Acknowledgement}  
We would like to thank H.\ Urbantke for his ongoing support, as
well as M.\ Kunzinger and H.\ Balasin for helpful discussions. 
This work was supported by the Austrian Science Foundation (FWF Project P-12023MAT).

\end{document}